# Multi-class Decoding of Attended Speaker Direction Using Electroencephalogram and Audio Spatial Spectrum

Yuanming Zhang, *Student Member, IEEE*, Jing Lu, *Member, IEEE*, Fei Chen, *Senior Member IEEE*, Haoliang Du, Xia Gao, Zhibin Lin

*Abstract*— Decoding the directional focus of an attended speaker from listeners' electroencephalogram (EEG) signals is essential for developing brain-computer interfaces to improve the quality of life for individuals with hearing impairment. Previous works have concentrated on binary directional focus decoding, i.e., determining whether the attended speaker is on the left or right side of the listener. However, a more precise decoding of the exact direction of the attended speaker is necessary for effective speech processing. Additionally, audio spatial information has not been effectively leveraged, resulting in suboptimal decoding results. In this paper, it is found that on the recently presented dataset with 14-class directional focus, models relying exclusively on EEG inputs exhibit significantly lower accuracy when decoding the directional focus in both *leave-one-subject-out* and *leave-one-trial-out* scenarios. By integrating audio spatial spectra with EEG features, the decoding accuracy can be effectively improved. The CNN, LSM-CNN, and Deformer models are employed to decode the directional focus from listeners' EEG signals and audio spatial spectra. The proposed Sp-EEG-Deformer model achieves notable 14-class decoding accuracies of 55.35% and 57.19% in *leave-one-subject-out* and *leave-one-trial-out* scenarios with a decision window of 1 second, respectively. Experiment results indicate increased decoding accuracy as the number of alternative directions reduces. These findings suggest the efficacy of our proposed dual modal directional focus decoding strategy.

## I. Introduction

The human brain can extract the speech of the attended speaker amid competing speakers and environmental noise [1], [2]. However, individuals with hearing impairments often struggle to understand speech, especially in noisy environments, necessitating the use of hearing aids [3], [4].

Modern hearing aids typically assume that the attended speaker is either the one producing the highest sound pressure level or the one located directly in front of the listener [5], [6]. However, these assumptions often prove inaccurate in real-world scenarios, as the desired speaker could be masked by interference or positioned significantly away from the listener. Determining the direction of the attended speaker from interfering speakers using only the signals captured by the hearing aids' microphones is infeasible without prior information, such as the pre-recorded speech of the attended speaker. Recent advancements in auditory attention decoding (AAD) [7], [8] have introduced a viable approach for decoding the direction of the attended speaker directly from brain signals, with electroencephalogram (EEG) being the most commonly used technique due to its non-invasive and convenient capture process [9], [10], [11], [12], [13], [14], [15].

The intricate nonlinear relationship between EEG signals and the attended audio poses a significant challenge for existing directional focus decoding methods. Rule-based methods, such as the filter bank common spatial patterns filter (FB-CSP) [16] and the Riemannian geometry-based classifier (RGC) [8], rely on second-order statistics to decode the attended direction. Deep neural network (DNN) approaches employ convolutional [7], recurrent [17], and self-attention [18] structures to directly extract the attended direction from EEG signals. The FB-CSP utilizes optimized filters to maximize the energy contrast of filtered signals between various directional focus classes [16]. The symmetric positive definite property enables the RGC to

†This work was supported by National Science Foundation of China with (Grant No. 12274221) and Postgraduate Research & Practice Innovation Program of Jiangsu Province (Grant No. KYCX24_0141). (Corresponding author: *Zhibin Lin.*).

Yuanming Zhang and Jing Lu are with Key Lab of Modern Acoustics, Nanjing University, Nanjing 210093, China, and also with the NJU-Horizon Intelligent Audio Lab, Horizon Robotics, Beijing 100094, China (e-mail: yuanming.zhang@smail.nju.edu.cn; lujing@nju.edu.cn)

Fei Chen is with the Department of Electrical and Electronic Engineering, Southern University of Science and Technology, Shenzhen 518055, China (e-mail: fchen@sustech.edu.cn).

Haoliang Du and Xia Gao are with the Department of Otolaryngology Head and Neck Surgery, Nanjing Drum Tower Hospital, Jiangsu Provincial Key Medical Discipline (Laboratory), Nanjing University, Nanjing 210008, China (e-mail: haoliangdu@163.com; gaoxiaent@163.com).

Zhibin Lin is with Key Lab of Modern Acoustics, Nanjing University, Nanjing 210093, China (e-mail: zblin@nju.edu.cn).

Color versions of one or more of the figures in this article are available online at http://ieeexplore.ieee.org

The EEG recording experiments were approved by the Institutional Review Board of Nanjing Drum Tower Hospital, with ethics approval number: 2022-065-09.

decode attended directions from the covariance matrices of EEG signals based on Riemannian distance [8]. These rule-based methods perform well with long EEG segments but degrade with shorter decision windows. A convolutional neural network (CNN) model [7] was proposed to directly extract patterns from EEG signals, achieving superior performance over rule-based methods and demonstrating greater robustness to changes in decision window lengths. The spatial-temporal attention network (STAnet) [18] and the spectro-spatial-temporal convolutional recurrent network [17] were developed to further enhance directional focus decoding accuracy. The STAnet dynamically assigns different weights to EEG channels. It employs a convolutional block attention module (CBAM) [19] to differentiate the contribution of different EEG channels, followed by a convolution layer and a multi-head attention (MHA) module [20] to extract temporal features. The CRN incorporates the azimuthal equidistant projection (AEP) method to transform EEG channels from their conventional 1D form into 2D images, facilitating better utilization of EEG sensor information. A convolutional recurrent network (CRN) is then applied to extract patterns in the 3D representation (2 spatial dimensions and 1 temporal dimension) of EEG signals. Recently, a learnable spatial mapping (LSM) module [21] is proposed to convert EEG channels into a 2D form, enhancing the performance of the CNN-based model.

Additionally, various self-attention-based models have been applied to facilitate different BCI tasks due to their high capacity for complex modeling [22], [23], [24], [25], [26], among which the Deformer model [22] has achieved superior classification accuracy in mental tasks, such as cognitive attention, driving fatigue, and workload detection, with a well-designed structure to capture both coarse- and fine-grained features. The information purification module in Deformer is also suitable for enhancing directional decoding accuracy.

Existing models predominantly focus on binary directional focus decoding, i.e., determining whether a listener is attending to the left or right side [7], [8], [16], [27], [28]. This limitation significantly hampers the practical application of decoding information, as a more precise angle of the attended speaker becomes crucial for subsequent speech processing. Although there has been research on a four-class directional focus decoding, its performance has not been fully validated in the *leave-one-subject-out* scenario [29]. Another recent presentation [30] also tried to decode the attended direction from 10 alternatives. However, they did not validate their results explicitly in a *leave-one-trial-out* or *leave-one-subject-out* manner. Their results were potentially overestimated due to trial-specific or subject-specific biases [31], [32]. Therefore, these datasets are not used in our paper.

Although existing models [7], [8], [16], [18], [21], [33] exhibit superior decoding accuracy, recent experiment results [31], [32], [34] highlight the significant impact of the cross-validation (CV) paradigm on decoding accuracy. These studies demonstrate that dividing a trial into training, validation, and test sets results in overestimated decoding accuracy, which is uncommon in traditional speech processing datasets.

Recently, an improved CV paradigm, known as *leave-one-out* (*LOO-CV*), was proposed to evaluate the decoding accuracy of an AAD model [7], [31]. In the *LOO-CV* paradigm, each trial is exclusively assigned to either the training, validation, or test set, preventing the model from leveraging trial-specific patterns to predict the labels of validation and test trials. Some researchers found that long-range temporal correlated (LRTC) components in EEG responses contribute to this phenomenon [32], [34], [35], [36], [37], [38]. However, the mechanism behind the long-range temporal correlation is beyond the scope of this paper. Nevertheless, *LOO-CV* is considered as a reasonable and necessary method for evaluating directional focus decoding models, as a BCI device would never capture EEG segments coming from its training set in practical use.

Decoding multi-class directional cues solely from EEG signals remains challenging. However, since BCI devices can easily incorporate both EEG electrodes and microphone units, a practical approach to improve multi-class directional focus decoding is to leverage the spatial cues embedded in audio signals. In this paper, a novel method that combines the spatial spectrum of multi-channel audio signals with EEG signals for directional focus decoding is proposed, evaluated on our recently released 14-class directional attention dataset. Our contributions are highlighted as follows.

- We elaborate on our newly released database, which includes 14 alternative speaker directions and enhances the capability of neural networks to decode the directional focus of attended speakers in a 14-class setting.
- We integrate the spatial spectrum with EEG features using a flexible module, plugged in the EEG-CNN, EEG-LSM-CNN, and EEG-Deformer models.
- We train and test our proposed models in challenging *leave-one-trial-out* and *leave-one-subject-out* scenarios. During testing, maximum effort to reduce trial-related bias that may lead to overestimated decoding accuracy is made.
- We use the EEG-Deformer model to decode the attended direction from listeners' EEG signals and compare its performance with that of the EEG-CNN and EEG-LSM-CNN models. Under the *LOO-CV* testing paradigm, the multi-class directional focus decoding accuracy was substantially lower than expected.

The remainder of this paper is organized as follows. Section II details our recently released multi-directional auditory attention dataset. Section III introduces the EEG-CNN, EEG-LSM-CNN, EEG-Deformer models, and their variants taking both EEG and spatial spectrum as inputs for decoding 14-class directional focus. This section also details the integration of the LSM module with the CNN model and the structure of the audio-EEG fusion convolution layer. Section IV provides details on training models and scientifically evaluating their decoding performance. Section V presents the experimental results, comparing the proposed models with existing approaches and conducting an ablation study to evaluate the efficacy of spatial-spectrum-fused directional focus decoding in various *LOO-CV* scenarios. This section also investigates the impact of decision window length and the number of alternative

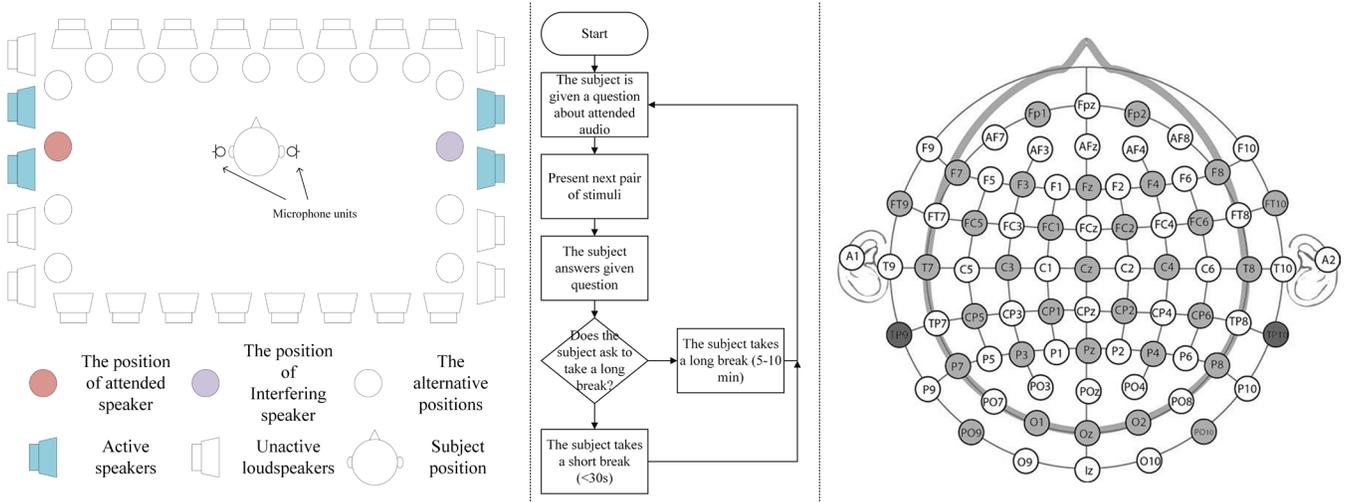

**Figure 1 A schematic diagram of the EEG experimental setup.** *Left*: The experimental setup. *Center*: The experimental flow chart. The directions of the competing speakers vary in each trial. *Right*: The electrode placement of the EEG recording device. The electrode distribution complies with the international 10-20 EEG electrode placement standard. Gray electrodes are used to capture EEG signals. The two black electrodes serve as reference channels. No electrodes are placed on white nodes.

directions on the decoding accuracy.

## II. THE NJU AUDITORY ATTENTION DECODING DATASET WITH MULTI-DIRECTIONAL CUES

### A. Experiment Setup

The NJU Auditory Attention Decoding Dataset was developed, encompassing multiple alternative directions of attended speakers [21], [39]. Our prior research focused on binary directional focus decoding [21]. This paper extends this research to explores the feasibility of multi-class directional focus decoding.

EEG data were collected from 28 young normal-hearing subjects as they listened to two competing speakers reproduced via a loudspeaker array. Subjects were instructed to focus on one specific speaker while ignoring the other. All participants provided formal written consent approved by the Institutional Review Board of Nanjing Drum Tower Hospital (ethics approval number: 2022-065-09) prior to the experiments and received financial compensation upon completion. Data from seven subjects were excluded from further analysis due to equipment malfunction.

EEG data were recorded using the 32-channel EMOTIV Epoc Flex Saline system at a sampling rate of 1024 Hz in a low-reverberation listening room. The electrode placement of the recording device is depicted in Figure 1. EEG signals were down-sampled to 128 Hz and wirelessly transmitted to the PC USB receiver. This device has been used to capture EEG signals, yielding similar results to other high-density EEG collecting devices [40], [41].

A schematic diagram of the experimental setup and procedure is shown in Figure 1. Auditory stimuli were reproduced through a loudspeaker array comprising thirty-two Genelec 8010 loudspeakers, all calibrated to the same sound pressure level. The auditory stimuli consisted of 32 Mandarin Chinese news programs narrated by several professional native-Mandarin-speaking news anchors. Some auditory stimuli were narrated by anchors of different genders, rendering this dataset unsuitable for investigating the impact of speaker gender on decoding accuracy. Silent or meaningless noisy segments were manually shortened. The stimuli were divided into 2-minute segments.

Each subject participated in 32 trials. In each trial, subjects were exposed to a pair of randomly selected and stimuli, with directions randomly and symmetrically drawn from 14 possible competing speaker directions, i.e., ±135°, ±120°, ±90°, ±60°, ±45°, ±30°, and ±15°. It is noted that our dataset originally consists of several trials in which the stimuli located at 0°, resulting in 15 possible competing speaker's directions. However, these trials, along with trials in which the competing directions are not symmetric, are not used in this paper. The total number of alternative directions is 14. Audio stimuli were presented at identical intensity through the loudspeaker using the vector-based amplitude panning (VBAP) algorithm [42], which panned the audio stimuli to their target directions. The loudspeakers were installed in an array measuring 2 meters in length and 1.5 meters in width. Notably, reproducing audio stimuli through loudspeakers and earphones yields similar quality of EEG signals and experimental outcomes [43], [44]. Subjects were instructed to attend to one of the two speakers while ignoring the other. Subjects were asked to close their eyes and remain still in the chair at the center of the loudspeaker array to minimize possible artifacts. Before each trial, subjects were given a question and instructed to find the answer in the attended audio stimuli. Subjects were allowed to take breaks after each trial to avoid listening fatigue. In total, there were 64 minutes (32 trials with 2-minute stimuli) of recordings per subject. However, our preprocessing removed some abnormal parts, therefore the available EEG recordings were shorter than 64 minutes per subject.

Additionally, an omnidirectional microphone was placed near each ear of the subject to capture the competing audio streams. It is important to note that determining the direction of the attended speaker using only these two-channel audio signals is infeasible due to existence of the competing speakers. However, the audio signals can be exploited as an effective auxiliary information to EEG for enhancing the decoding performance.

### B. EEG Signal Preprocessing

EEG preprocessing was performed using the EEGLAB toolbox [45]. The EEG signals were filtered through a bandpass filter between 1 Hz and 32 Hz, consistent with most previous AAD tasks [16], [27], [44]. EEG signals were decomposed through independent component analysis (ICA), and components containing eye, heart, or muscle artifacts were removed. As instructed by EEGLAB, the EEG signal underwent interpolation to restore the excluded channels.

### C. Audio Signal Preprocessing

Audio recordings with an original sampling rate $f_s = 44.1$ kHz were down-sampled to 8 kHz. The down-sampled audio signals were then transformed into the time-frequency domain via short-time Fourier transform (STFT) with $F$ frequency bins and $N$ time frames, with each time-frequency bin expressed as $Y_l(f,n)$ with $l$ the microphone unit index, $f$ the frequency index, and $n$ the time frame index. Subsequently, the minimum variance distortionless response (MVDR) beamformer [46] was applied to extract the spatial spectrum, which can be represented as

$$P_{mvdr}(\theta) = \frac{1}{F}\sum_{f=0}^{F-1} \frac{1}{\mathbf{g}(f,\theta)^T \mathbf{R}(f)^{-1} \mathbf{g}^*(f,\theta)}, \quad (1)$$

where $\mathbf{g}(f,\theta)$ is the steering vector, $\mathbf{R}(f)$ is the correlation matrix of the microphone array defined as

$$\mathbf{R}(f) = \begin{bmatrix} \mathbf{y}_{1,f}^T \mathbf{y}_{1,f}^* & \mathbf{y}_{1,f}^T \mathbf{y}_{2,f}^* \\ \mathbf{y}_{2,f}^T \mathbf{y}_{1,f}^* & \mathbf{y}_{2,f}^T \mathbf{y}_{2,f}^* \end{bmatrix}, \quad (2)$$

with $\mathbf{y}_{l,f}$ a column vector written as

$$\mathbf{y}_{l,f} = [Y_l(f,0), Y_l(f,1),...,Y_l(f,N-1)]^T, l=1,...,L. \quad (3)$$

The left microphone unit is designated as the reference point. The angle $\theta$ begins at $-90°$ on the left-hand side and extends to $90°$ on the right-hand side. The steering vector is then simplified to

$$\begin{aligned}\mathbf{g}(f,\theta) &= [g_1(f,\theta), g_2(f,\theta)]^T \\ &= [1, \exp(-2\pi i f \sin(\theta) d / c)]^T\end{aligned}. \quad (4)$$

The first ($l = 1$) and second microphone ($l = 2$) are placed near the left and right ear, respectively.

## III. PROPOSED METHODS

### A. Overview of Models

Figure 2 depicts the architecture of the models used to decode the direction of attended speakers. Figure 2 (a-1) illustrates the general pipeline of the EEG-based directional focus decoding. A DNN model is trained using EEG signals to predict the most probable direction of the attended speaker. Figure 2 (b-1) presents our proposed audio-EEG directional focus decoding method, in which a spatial spectrum is also provided as the second input of a DNN classifier to boost the directional focus decoding. The remaining sections of the first row of Figure 2 detail the structures of the EEG-CNN [7], EEG-LSM-CNN [21], and EEG-Deformer [22] models. While the structures of Sp-EEG-CNN, Sp-EEG-LSM-CNN, and Sp-EEG-Deformer models, which integrates spatial spectrum information with EEG features to improve decoding accuracy, are exhibited in the second row.

Let $\mathcal{X} \in \mathbb{R}^{C \times T}$ be the input EEG with $C$ and $T$ representing the total number of EEG channels and time sample points respectively. The EEG-CNN model, shown in Figure 2 (a-2), directly takes $\mathcal{X}$ as input and performs convolution along the channel dimension. An average pooling layer condenses the intermediate features, followed by fully connected layers to output the predicted class labels.

Let $\mathbf{p}_s \in \mathbb{R}^{N_\theta}$ be the spatial spectrum with $N_\theta$ the number of spatial sampling points of the spectrum. A fusion block is inserted before the first convolution block in the Sp-EEG-CNN model (Figure 2 (b-2)) to leverage both EEG and audio spatial information. The fusion block first transforms the spatial spectrum into high-dimensional patterns and then concatenates them with EEG features to form a new tensor. The subsequent convolution block uses the fused tensor as input to determine the attended directions.

Figure 2 (a-3) and (b-3) depict the structure of the proposed EEG-LSM-CNN and Sp-EEG-LSM-CNN models, respectively. At the start of the model, a learnable spatial mapping (LSM) module is added, designed to utilize the spatial information of EEG sensors, transforming the EEG channels from 1D into a 2D discretized plane. The output of the LSM module is $\mathcal{Z} \in \mathbb{R}^{C_1 \times C_2 \times T}$, where $C_1$ and $C_2$ are the hyperparameters that specify the size of the grids used to describe the target 2D plane. A detailed explanation of the LSM module can be found in the next subsection and our previous research [21], [33]. It is noted that the LSM module leverages the spatial information of EEG electrodes (see Figure 1) but not the spatial information of audio signals, and the Sp-EEG-LSM-CNN model fuses the mapped EEG features with the spatial spectrum instead of the raw EEG signals.

The remaining layers in EEG-LSM-CNN resemble those of the conventional network [7], with the only difference being the use of 3D convolution blocks (instead of 2D convolution), followed by an average pooling layer to condense the time dimension. The network uses a flatten layer and Fully Connected (FC) layers to generate the output, which is the attended direction.

The EEG-Deformer model (Figure 2 (a-4)) [22], which demonstrates superior performance in other BCI fields, is a new variant of transformer [20], [23], [24]. The EEG-Deformer is adopted to leverage the audio spatial spectrum by inserting a fusion block after the first EEG encoder. The remainder of Sp-EEG-Deformer is the same as the original.

### B. The Learnable Spatial Mapping Module

Current data-driven methods primarily implement image or speech processing models for directional focus decoding [7], [18]. However, these methods fail to effectively leverage the spatial distribution of EEG electrodes, leading to suboptimal performance. Although EEG sensors (electrodes) are spatially distributed in 3D space, EEG signals are conventionally represented as multi-channel signals with additional descriptive text for each channel. Unfortunately, these texts cannot be directly utilized by DNN models. Therefore, it is crucial to design more effective mechanisms to exploit the spatial distribution of EEG sensors.

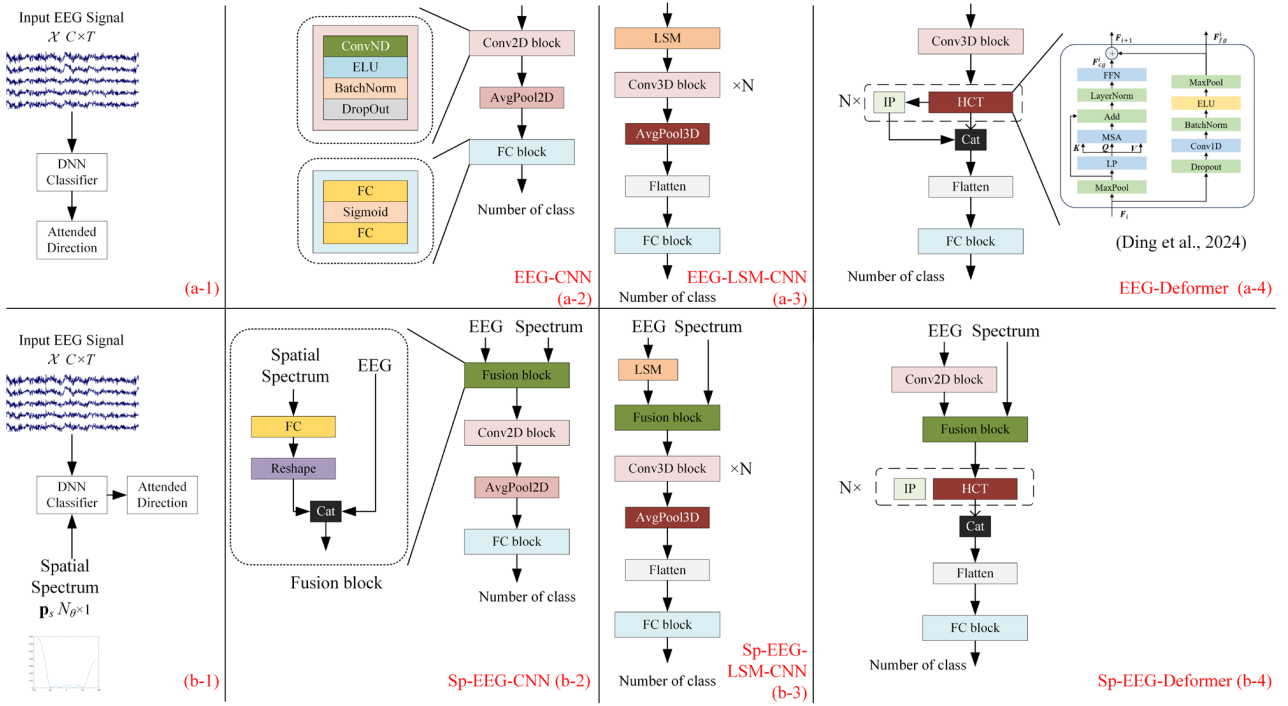

**Figure 2 The pipeline of DNN-based multi-class directional focus decoding and the structures of various DNN classifiers.** (a/b-1 Pipelines of directional focus decoding with and without spatial information. (a-2~4) Structures of EEG-CNN, EEG-LSM-CNN, and EEG-Deformer. These models utilize only EEG signals to capture the directions of the attended speakers. (b-2~4) Structures of Sp-EEG-CNN, Sp-EEG-LSM-CNN, and Sp-EEG-Deformer. These models leverage both EEG signals and spatial information to enhance directional focus decoding performance.

An intuitive method for reconstructing spatial information involves transforming the 1D distributed EEG channels onto a 2D uniformly discretized plane based on their original positions. However, our experimental results using this approach exhibited a decline in performance. One potential explanation for this outcome is that the direct mapping significantly deviates from the actual distribution, thereby failing to adequately leverage the available spatial information.

The proposed LSM module is illustrated in Figure 2 (a-3). $C \times 1 \times 1$ convolution filters are applied to process EEG signals $\mathcal{X}$, and the number of convolution filters equals the number of channels on the 2D plane. These convolution filters have flexible weights updated during backpropagation and are shifted across the temporal dimension to linearly combine all EEG channels, resulting in new virtual channels. The convolution layer contains a total of $C_1 C_2$ kernels, matching the number of virtual channels. Subsequently, a batch normalization layer is applied. The output of the convolution layer can be expressed as

$$\widehat{\mathcal{Z}}(k,t) = \mathrm{BN}(\sum_{c=1}^{C} \mathcal{X}(c,t)\mathcal{G}_k(c)) \quad (5)$$

where $\mathcal{G}_k$, $k = 1,2,\ldots,C_1 C_2$, is the $k$-th kernel of the convolution layer in LSM. The output $\widehat{\mathcal{Z}} \in \mathbb{R}^{C_1 C_2 \times T}$ is an intermediate tensor, which is then rearranged into $\mathcal{Z} \in \mathbb{R}^{C_1 \times C_2 \times T}$ as

$$\mathcal{Z}(c_1, c_2, t) = \widehat{\mathcal{Z}}(c_2 + C_2 \times (c_1 - 1), t) \quad (6)$$

with $t$ the temporal index, $c_1$ and $c_2$ the indices of the 2D plane, respectively.

In addition to incorporating the connectivity between neighboring channels, our proposed LSM module reorganizes the spatial layout of EEG channels. This feature allows subsequent convolution layers to emphasize the intrinsic relationships among channels positioned within the 2D plane. This behavior mimics the simultaneous activation patterns observed in human brains, where neurons tend to activate in conjunction with neighboring neurons [47]. It is important to note that in the context of LSM, the term "spatial" refers to the 2D plane learned by the module, as opposed to the conventional 1D arrangement of EEG recordings.

### C. Fusion of spatial spectrum and EEG features

The dotted box in Figure 2 (b-2) depicts the structure of the fusion block, which takes both spatial spectrum and EEG feature as inputs. An FC layer is applied to transform the spatial spectrum into a high-dimensional embedding, which is then reshaped to match the dimensions of the EEG features. The reshaped spatial spectrum embedding is concatenated with the EEG feature to form the output of the fusion block. This approach facilitates the neural network utilizing spatial cues of the competing speakers.

Specifically, let $\mathcal{Z}$ be the intermediate EEG feature, e.g., the output of the LSM module. The fused feature $\widetilde{\mathcal{Z}}$ can be represented as

$$\begin{aligned}\hat{\mathbf{p}} &= \text{FC}(\mathbf{p}_s) \\ \mathcal{P} &= \text{reshape}(\mathbf{p}, C_1, C_2, 1) \\ \widetilde{\mathcal{Z}} &= \text{concat}(\mathcal{Z}, \mathcal{P})\end{aligned} \quad (7)$$

## IV. DATA PREPERATION & TRAINING

### A. Cross-Validation

It has been revealed that trial-specific features in EEG signals severely impacts DNN-based directional focus decoding models, resulting in overestimated decoding accuracy [31], [32], [34], [35], [36], [37], [38]. Speaker-related and audio-content-related features may also be captured by DNN models, necessitating carefully designed experiments to mitigate these effects. The participants are required to close their eyes during audio playback to minimize visual bias. However, eye movement components cannot be fully controlled during EEG experiments nor removed by preprocessing algorithms, resulting in unavoidable visual bias in our dataset.

In our experiments, we make our best effort to remove potential biases. Figure 3 demonstrates the employed *LOO-CV* paradigms. Figure 3 (a) illustrates the *leave-one-trial-out (LOTO)* scenario, where the training, validation, and test trials never overlap. Additionally, the DNN decoder is trained on the training trials of all subjects and evaluated on the held-out trials. That is to say, the *LOTO* decoder depends on specific subject groups.

Figure 3 (b) depicts the *leave-one-subject-out (LOSO)* scenario. In this scenario, trials from one subject are used to test the models, while trials from other subjects are used to train the DNN decoder. In other words, the DNN decoder never sees any trials from the test subject. Thus, the decoder cannot leverage subject-specific patterns to achieve higher decoding accuracy, making the *leave-one-subject-out* scenario more generalized and more challenging.

Figure 3 (c) demonstrates the *leave-one-moment+trial-out (LOMTO)* condition. In this paradigm, trials adjacent to the validation or test trials (white trials in the figure) are removed. Figure 3 (d) and (e) depict the *leave-one-class+trial-out (LOCTO)* and *leave-one-audio+trial-out (LOATO)* conditions, respectively. The *LOCTO* datasets follow these rules:

- Test trials from one subject may have only one unique class label. Trials from different subjects must have distinct class labels.
- If a trial is included in one dataset, trials possessing the same class label are also included in that dataset.
- The total number of class labels of one dataset is equal to the number of alternative attention directions, which is 14 in our dataset.

The *LOATO* datasets apply similar restraints to the attended audio:

- Test trials from one subject may have only one unique attended audio.
- If a trial is included in one dataset, trials with the same attended audio are also included in that dataset.
- The total number of class labels of one dataset is equal to the number of alternative attention directions, which is 14 in our dataset.

All cross-validation and training procedures are implemented using PyTorch [48]. The Adam optimizer is used during the

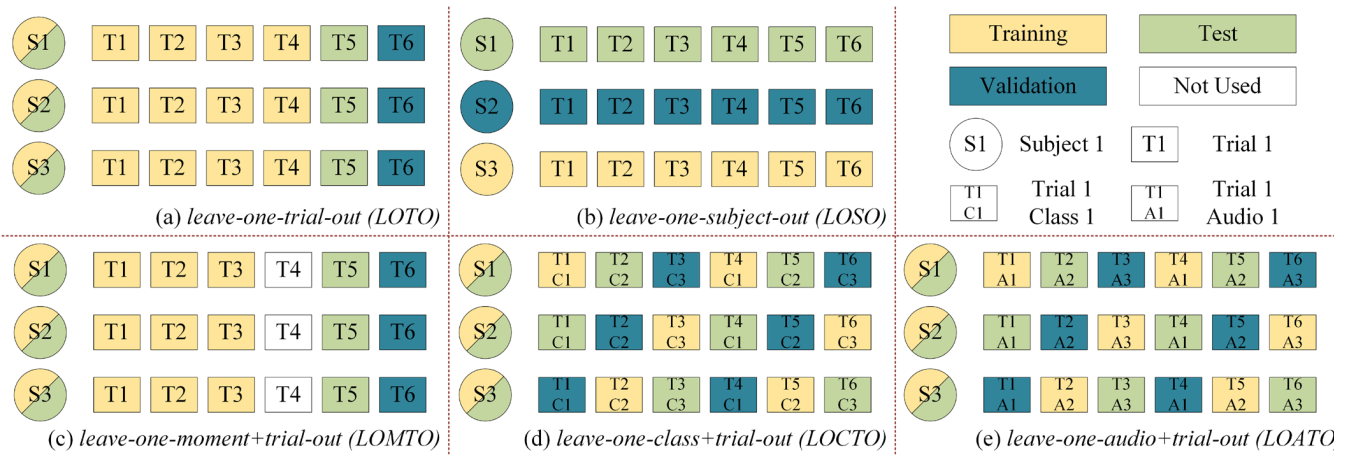

(a) *leave-one-trial-out (LOTO)*  (b) *leave-one-subject-out (LOSO)*  (c) *leave-one-moment+trial-out (LOMTO)*  (d) *leave-one-class+trial-out (LOCTO)*  (e) *leave-one-audio+trial-out (LOATO)* training, and a learning rate drop-down policy is applied to improve convergence. The L2 regularization is applied to further alleviate overfit.

**Figure 3 A schematic diagram of the *leave-one-out* cross-validation paradigms.** (a) In the *LOTO* scenario, some trials are held out to comprise the validation and test sets, and the remaining trials form the training set. (b) In *LOSO* scenario, data from held-out subjects form the validation and test sets, and data from the remaining subjects form the training set. (c) In the *LOMTO* scenario, the training and evaluation sets consist of nonadjacent trials. (d) In the *LOCTO* scenario, validation and test trials from different subjects have unique class labels. (e) In the *LOATO* scenario, validation and test trials from different subjects have unique attended audios. Note that this diagram is for demonstration only, the actual numbers of subjects, trials, classes, and attended audios depend on the dataset.

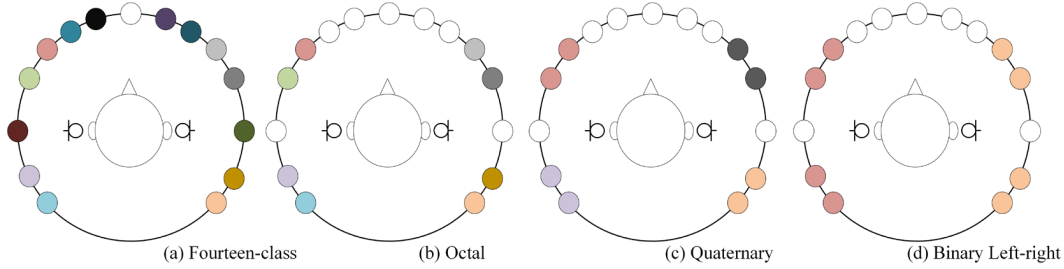

**Figure 4 A schematic diagram of the labeling paradigm of our proposed dataset.** Various colors indicate various class labels. Non-colored directions are not used.

### B. Trial Labeling

Figure 4 illustrates four distinct trial labeling paradigms. In the first paradigm, trials where the attended speaker is located in front of the listener are excluded. The remaining trials are labeled as their directions of attended speakers. In the second paradigm, trials without front-rear counterparts are excluded from the dataset, and the retained trials are labeled as their original attended directions, resulting in eight distinct classes. In the third paradigm, trials retained in the second paradigm are labeled according to the quadrant where they live, leading to a 4-class directional focus decoding dataset. Finally, the fourth paradigm labels trials as left-right symmetric pairs.

The proposed labeling paradigms facilitate the investigation of impact of number of alternative directions on the performance of directional focus decoding, which is discussed in next section.

### C. Spatial Spectrum Computation

The spatial spectrum for each trial was precomputed following Equation (1), and all segments of a single trial were linked to the same spatial spectrum.

### D. Model performance evaluation

The decoding accuracy is used to evaluate the performance of a DNN directional focus decoder. The decoding accuracy can be expressed as

$$\text{acc} = \frac{1}{N_{class}} \sum_{n=1}^{N_{class}} \text{acc}_n = \frac{1}{N_d} \sum_{n=1}^{N_{class}} \frac{M_{correct,n}}{M_n} \quad (8)$$

with $N_{class}$ the number of alternative directions, and $M_{correct,n}$, $M_n$ the number of correctly classified samples and the number of total samples in class $n$, respectively.

Wilcoxon sign-rank test is used to compare our decoding results against the chance level. Specifically, any model solely utilizing EEG signals is compared against a random guess decoder. The chance level of the random decoder is set to $1/N_{class}$. Random guess samples are generated from a binomial distribution with the specified chance level. The number of binomial-distributed samples is equal to that in test sets. And the null hypothesis for a model taking dual modal inputs (EEG and spatial spectrum) is that its decoding accuracy is not significantly higher than its twin taking solely EEG as input, and a binary random guess decoder. For example, only if the Sp-EEG-CNN significantly surpasses the EEG-CNN and a binary decoder, the null hypothesis is rejected.

Additionally, bootstrapping is used to obtain significant levels of the sign-rank test results. 1 million times bootstrapping is applied, and the 95-percentile of the bootstrapping results (a.k.a. p-values) is considered as the final significant level.

## V. RESULTS & DISCUSSION

### A. 14-class directional focus decoding accuracy

Table 1 demonstrates the decoding accuracy of the EEG-CNN, EEG-LSM-CNN, and EEG-Deformer models, with their variants taking both EEG and spatial spectrum as inputs, on our proposed 14-class AAD dataset in the *LOTO* and *LOSO* paradigms with various window lengths. Although some models which solely utilize EEG signals surpass the chance level of a 14-class decoder, the decoding performance of these models is quite below our expectation, as shown in Table 1. Surprisingly, the EEG-CNN model achieves the highest decoding accuracy among all models which only take EEG as input. Its decoding accuracy is 13.05% and 10.25% on the *LOTO* condition with a window length of 10 seconds and 1 second, respectively. Significant difference is found between the two window lengths ($p < 0.001$). The third-ranked model is EEG-LSM-CNN. It achieves a decoding accuracy of 9.83% on the *LOSO* condition with a window length of 10 seconds.

All models achieve better decoding performance by integrating audio spatial spectra. Although the Sp-EEG-CNN model did not surpass the chance level on the *LOSO*-1s condition, it again achieves the highest decoding accuracy on the *LOTO*-1s condition, as shown in Table 1. Other two models also benefit from the fusion of spatial spectra and show significantly higher decoding performance compared to a binary random decoder.

All models solely utilizing EEG features attain a higher decoding accuracy on a longer decision window length, which is consistent with previous studies [7], [16], [49]. However, they achieve deteriorated performance with longer EEG samples when incorporating dual modal inputs. The Sp-EEG-CNN model dramatically deteriorates when the decision window length increases. A possible explanation is that the single-layer structure of the CNN model [7] constructs a limited temporal receptive field, resulting in degraded feature recognition capability. The duplicated spatial spectrum features in temporal dimension may hamper the learning of fluctuating features, resulting in degraded performance. Due to its performance deterioration, the CNN model is excluded from further analysis.

**Table 1 Decoding accuracy of EEG-CNN, EEG-LSM-CNN, EEG-Deformer models and their variant ones in the *LOSO* and *LOTO* scenarios.** The number of alternative attended direction is set to 14, the maximum of our dataset. "EEG" denotes a model solely taking EEG signals as input. And "Sp+EEG" denotes a model taking both EEG signals and spatial spectrums as input. Asterisk(s) (*) indicates the significant level of Wilcoxon sign rank test. Details of statistical tests are elaborated in III.D.

| Model | CV paradigm | *LOSO* | | *LOTO* | |
|---|---|---|---|---|---|
| | Sample length | 1 second | 10 seconds | 1 second | 10 seconds |
| EEG-CNN | | 8.68%*** | 9.07%* | 10.25%*** | 13.05%*** |
| Sp-EEG-CNN | | 47.52% | 22.92% | 58.33%*** | 21.76% |
| EEG-LSM-CNN | | 8.98%*** | 9.83%** | 6.14% | 7.32% |
| Sp-EEG-LSM-CNN | | 53.50%*** | 53.12% | 55.69%*** | 50.05% |
| EEG-Deformer | | 8.32%*** | 9.44%** | 8.02%** | 9.22% |
| Sp-EEG-Deformer | | 55.35%*** | 52.99% | 57.19%*** | 51.61% |

### B. The impact of number of alternative directions

In addition to the 14-class labeling paradigm, we propose three more paradigms with fewer alternative directions to validate the efficacy of our EEG dataset, as demonstrated in Figure 4.

Table 2 presents the decoding accuracy of EEG-LSM-CNN and EEG-Deformer models and their variants in the binary (2-class), quaternary (4-class), and octal (8-class) directional focus decoding scenarios. It is shown that all models achieve higher decoding accuracies as the number of alternative directions decreases Specifically, the EEG-Deformer model obtains *LOSO* decoding accuracies of 62.42%, 31.14%, and 16.10% in the binary, quaternary, and octal directional focus decoding experiments, respectively. The EEG-LSM-CNN model achieves *LOSO* decoding accuracies of 67.27%, 33.64%, 18.83%, respectively. Significant differences between the decoding accuracies of EEG-LSM-CNN, EEG-Deformer and the random guess decoder are found. Therefore, the feasibility of multi-class directional focus decoding solely based on listeners' EEG signals and the efficacy of our dataset is validated. Additionally, experimental results suggest that auditory-evoked EEG responses contain effective but limited information about the attended direction of the listeners, leading to degraded performance of 8-class and 14-class directional focus decoding.

Experimental results suggest that, rather than simply reducing the alternative directions to the locations of competing speakers, fusing and leveraging the spatial spectrum and EEG features more effectively decodes the attended direction. Specifically, the Sp-EEG-Deformer model achieves decoding accuracies of 72.27%, 69.39%, and 53.56% in the binary, quaternary, and octal *LOSO* scenarios. Additionally, the number of alternative directions also negatively impacts the decoding performance of the Sp-EEG-Deformer model when the spatial spectrum is integrated. This result suggests that, instead of a simple concatenation, a potential integration of EEG features and spectrum features exists in the deep structure of the network. The Sp-EEG-LSM-CNN model achieves similar results, with decoding accuracies of 60.45%, 62.58% and 55.61%, in the binary, quaternary, and octal *LOSO* scenarios, respectively.

### C. The impact of spatial spectrum

Although the advantage of leveraging audio spatial cues in directional focus decoding has been demonstrated in previous subsections, as well as Table 1 and Table 3, it is necessary to further discuss the impact of spatial spectrum fusion to provide extra insight into the fusion of spatial spectrum and EEG signals.

As depicted in Table 1 and Table 3, both Sp-EEG-LSM-CNN and Sp-EEG-Deformer models achieve significantly higher decoding accuracy by integrating the audio spatial spectrum and EEG features. Additionally, they achieve decoding accuracy above the chance level of a binary decoder, suggesting the effective integration and utilization of both features, as relying solely on audio features results in a binary random guess decoder.

**Table 2 Decoding accuracy of EEG-LSM-CNN, EEG-Deformer models and their variant ones with various number of alternative directions.** The decision window length is 10 seconds. "EEG" denotes a model solely taking EEG signals as input. And "Sp+EEG" denotes a model taking both EEG signals and spatial spectrums as input. Asterisk(s) (*) indicates the significant level of Wilcoxon sign rank test. Details of statistical tests are elaborated in III.D.

| CV-paradigm / Model | Binary (2-class) | | Quaternary (4-class) | | Octal (8-class) | |
|---|---|---|---|---|---|---|
| | *LOSO* | *LOTO* | *LOSO* | *LOTO* | *LOSO* | *LOTO* |
| EEG-LSM-CNN | 67.27%*** | 57.23%** | 33.64%*** | 34.44%*** | 18.83%*** | 10.19% |
| Sp-EEG-LSM-CNN | 60.45% | 64.01%** | 62.58% | 57.49%** | 55.61%** | 55.37% |
| EEG-Deformer | 62.42%*** | 55.07% | 31.14%*** | 28.35% | 16.10%* | 14.15% |
| Sp-EEG-Deformer | 72.27%*** | 64.02%*** | 69.39%*** | 56.83%** | 53.56% | 35.66% |

The microphone array used in our experiment contains two units, one placed near the left ear and another near the right ear. Our dataset includes some attended directions that are front-rear symmetric, such as 60° versus 120°, and -45° versus -135°. Traditionally, a two-element microphone array cannot distinguish whether the sound source is coming from the front or rear. However, an array placed near the head provides not only the left/right spatial cues but also informs the DNN model of the front/rear information. (If not, the DNN model tends to become a four-class random guess decoder.) A convincing explanation is that the presence of the listener's head affects the time delay of audio signals, resulting in divergent spatial spectra. This variance is then recognized by the DNN model and reflected in the contrasting audio spatial features.

As depicted in Figure 4, only trials with attended direction ±135°, ±120°, ±60°, and ±45° are retained in the dataset and train DNN models on it to validate their capability to decode front-rear spatial cues. The proposed Sp-EEG-Deformer achieves a decoding accuracy of 69.39% in the 4-class scenario, exhibiting a strong capability to distinguish the spatial cues hidden in the audio spatial spectrum.

### D. The impact of unseen subject

An ideal auditory BCI device operates on users excluded from its training set. Additionally, a *LOTO* decoder tends to learn LRTC components, resulting in overestimated decoding accuracies [31], [32], [34], making *LOSO* decoders more valuable. Such a LOSO decoder cannot learn LRTC components which exist in adjacent trials of the same subject, since the validation and test trials never come from training subjects. A schematic demonstration is shown in Figure 3.

Extensive experiments confirm a slight degradation of a 14-class directional focus decoder in the *LOSO* scenario, consistent with previous reports [16], [49]. However, both EEG-LSM-CNN and EEG-Deformer achieve higher decoding accuracy in the 2-class, 4-class, and 8-class *LOSO* decoding experiments, compared to their *LOTO* ones. This conclusion still holds for Sp-EEG-Deformer. These results imply the existence of eye-gaze-related components, which is consistent with previous findings [31], [32]. Specifically, DNN models might learn eye-gaze- or eye-movement-related components to establish a link from the eye-related components to the direction of the attended speaker. This eye-related bias gets weaker as the number of alternative directions increase. Although participants were asked to close their eyes, they were not told to avoid looking at the direction of the attended speaker with their closed eyes. Besides, autonomous and uncontrolled eye gazing behaviors may still exist even if the subjects were asked to remain their eye fixed during experiment trials. If subjects move their eyes to focus on the direction of attended speaker in every trial, one is expected to be capable of training a DNN model to capture such eye-movement-related components to classify trials into two clusters, namely the "left" and "right", representing the eye-focused direction. However, as the number of alternative directions increase, it is much harder to tell which left class (e.g., left-front and left-rear) the listener is 'attending' to if his/her eyes focus on the left side. Generally, experiment results imply the possibility of existence of eye-related bias in EEG-AAD data. However, further investigation of eye-related bias is far beyond the scope of this paper.

### E. The impact of unseen trials with additional constraints

To the best of our knowledge, previous research has not fully investigated the impact of the *LOTO* scenario with additional constraint conditions. Therefore, extensive experiments are conducted to identify any potential bias. Specifically, three *LOTO* datasets are constructed with additional various constraints. A schematic description of the three datasets is shown in Figure 3.

As shown in Table 3, the Sp-EEG-LSM-CNN and Sp-EEG-Deformer models achieves similar results on the *LOMTO* dataset, compared to the *LOTO* one. Unlike binary directional focus decoding, adjacent trials in our 14-class dataset inherently possess distinct attention labels. Therefore, it is not problematic for our proposed models to accommodate the non-adjacent dataset.

The *LOATO* dataset poses a greater challenge for DNN models. Table 3 demonstrate a moderate degradation in the 14-class decoding accuracy of our proposed DNN models. Surprisingly, the Sp-EEG-LSM-CNN model obtains higher decoding accuracy on the held-audio-out class dataset with a decision window length of 10 seconds. This decline is attributed to the unseen attended *audio* in the validation and test sets and suggests potential overfitting of DNN models on the audio-dependent features.

**Table 3 Decoding accuracy of EEG-LSM-CNN, EEG-Deformer models and their variant ones in the *LOMTO, LOATO,* and *LOCTO* scenarios with various number of alternative directions.** "EEG" denotes a model solely taking EEG signals as input. And "Sp+EEG" denotes a model taking both EEG signals and spatial spectrums as input. Asterisk(s) (*) indicates the significant level of Wilcoxon sign rank test. Details of statistical tests are elaborated in III.D.

| Model | CV paradigm | *LOTO* | | *LOMTO* | | *LOATO* | | *LOCTO* | |
|---|---|---|---|---|---|---|---|---|---|
| | Sample length | 1 second | 10 seconds | 1 second | 10 seconds | 1 second | 10 seconds | 1 second | 10 seconds |
| EEG-LSM-CNN | | 6.14% | 7.32% | 6.07% | 6.64% | 7.26% | 7.90% | 4.17% | 3.57% |
| Sp-EEG-LSM-CNN | | 55.69%*** | 50.05% | 55.46%*** | 48.56% | 52.67% | 53.62% | 46.60% | 53.33% |
| EEG-Deformer | | 8.02%** | 9.22% | 6.58% | 9.36% | 7.36% | 8.70% | 5.82% | 8.85% |
| Sp-EEG-Deformer | | 57.19%*** | 51.61% | 57.10%*** | 51.78% | 55.53%*** | 52.08% | 47.66% | 55.11% |

Finally, the proposed models are trained on the *LOCTO* dataset. The *LOCTO* dataset introduces a new limitation for DNN models by excluding training trials with the same attended *direction* as the evaluation trials. Hence, the model cannot utilize features from the same-class training trials to obtain a higher decoding accuracy. Additionally, this method potentially eliminates the LRTC component bias, as the model never learns any LRTC features that correspond to the validation and test trials. It is evident that both models obtain significantly deteriorated decoding accuracy on the held-out class dataset.

In general, new limitations to *LOTO* datasets pose significant challenges on DNN models. Nevertheless, it is necessary to conduct such experiments to fully evaluate the generalization capability and robustness of a data-driven directional focus decoding model. As our EEG dataset is not designed for these experiment conditions, e.g., unseen attended audio, it is impossible to construct a *leave-one-audio+subject-out* dataset to evaluate the performance of DNN decoders. Carefully designed EEG datasets are desired to meet the requirements for conducting such experiments.

## VI. CONCLUSIONS

This paper presents our newly recorded NJU auditory attention decoding dataset. To the best of our knowledge, this is the first EEG recording dataset featuring 14 alternative directions for attended speakers. Using this dataset, the feasibility of 14-class decoding of directional focus is validated, which offers more effective information for subsequent speech processing compared to the commonly investigated binary decoding. Furthermore, we propose a novel approach integrating audio spatial features with EEG features to enhance the performance of 14-class directional focus decoding. This approach utilized a fusion block to combine the two feature modalities to determine the attended directions. Extensive experiments are conducted using the *leave-one-out* cross-validation paradigm, known for its realistic and stringent approach for testing directional focus decoders, to assess the feasibility of our proposed method. Specifically, it is found that the 14-class decoding accuracy to be quite low, close to the chance level. However, the experimental results show that the proposed Sp-EEG-Deformer model can achieve up to 57.19% average accuracy. The Sp-EEG-LSM-CNN model also achieves an average decoding accuracy of 55.69%. Statistical analysis shows a significant difference between the accuracy of the Sp-EEG-LSM-CNN and Sp-EEG-Deformer models and the chance level of binary classification, indicating that our proposed methods successfully leverage both audio and EEG signals to decode the directions of attended speakers from 14 alternative directions. Furthermore, the robustness of our models in the *leave-one-audio+trial-out, leave-one-moment+trial-out, and the leave-one-class+trial-out* scenarios are validated. Results indicate severe deterioration when one class is unseen during training. In contrast, the long-range temporal correlated component has little impact on the decoding results, as participants are always required to attend to different directions in adjacent trials. Our future research will focus on more effective model which better combines audio spatial cues and EEG features.